\documentclass[article]{aa} 
\usepackage{graphicx}
\begin{document}

 \title{Unveiling the structure of the planetary nebula M~2-48:}

   \subtitle{Kinematics and physical conditions.}

\author{L. L\'opez-Mart\'\i n\inst{1,2}, J. A. L\'opez \inst{3}, C. Esteban \inst{1}, R. V\'azquez\inst{3}, A. Raga\inst{4}, J. M. Torrelles\inst{5}, L. F. Miranda  \inst{6}, J. Meaburn \inst{7}, and L. Olgu\'\i n \inst{8}}
  
\offprints{Luis L\'opez-Mart\'\i n} 

\institute{Instituto de Astrof\'\i sica de Canarias, c/V\'\i a L\'actea s/n, 38200 La Laguna, Tenerife, Spain\\
	email: luislm,cel@ll.iac.es
		\and
Observatoire de Paris, DEMIRM, UMR 8540 du CNRS, 61 Avenue de l'Observatoire, 75014 Paris, France\\
              email: luis.lopezmartin@obspm.fr
              \and
Instituto de Astronom\'\i a, UNAM, Km 103, Carretera Tijuana-Ensenada, 22860 Ensenada, B. C., Mexico \\
              email: jal,vazquez@astrosen.unam.mx
	      \and
Instituto de Ciencias Nucleares, Circuito de la Investigaci\'on Cient\'\i fica, 04510 Cd. Universitaria, D. F., Mexico\\
		email:raga@astroscu.unam.mx
		\and
Institut d'Estudis Espacials de Catalunya and Instituto de Ciencias del Espacio (CSIC), Edifici Nexus, c/ Gran Capit\'a 2-4, 08034, Barcelona, Spain\\
		email:torrelles@ieec.fcr.es
		\and
Instituto de Astrof\'\i sica de Andaluc\'\i a, CSIC, c/ Sancho Panza s/n, 18008 Granada, Spain\\
		email:lfm@iaa.es
		\and
Astronomy Department, University of Manchester, Jodrell Bank, Macclesfield, Cheshire SK11 9DL, UK
 \and
  Instituto de Astronom\'\i a, UNAM, Circuito de la Investigaci\'on Cient\'\i fica, 04510 Cd. Universitaria, D. F., Mexico\\
              email: lorenzo@astroscu.unam.mx      }
           
   \date{Received ------------; accepted ----------}

\abstract{

The kinematics and physical conditions of the bipolar
planetary nebula M 2-48 are analysed from high and low dispersion
long-slit spectra.  Previous CCD narrow-band optical observations have 
suggested that this nebula is mainly formed by a pair of symmetric bow-shocks,
 an off-center semi-circular shell, and an internal bipolar structure.
The bipolar outflow has  a complex
structure, characterised by a series of shocked
regions located between the bright core and the polar tips. There
is an apparent kinematic discontinuity between the bright bipolar core
and the outer regions. The fragmented ring around the bright bipolar region
presents  a low expansion velocity and could be associated
to ejection in the AGB-PN  transition phase, although its
nature remains unclear.
The chemical abundances of the central region are derived,
showing that M~2-48 is a Type~I planetary nebula (PN).

{\bf keywords:} ISM: jets and outflows --- echelle spectroscopy ---
Hydrodynamics --- Shock waves --- planetary nebulae : individual: M
2-48
}
\authorrunning{ L\'opez-Mart\'\i n et al.}
\titlerunning{Unveiling the structure of the planetary nebula M~2-48}
 \maketitle

\section{Introduction}

\noindent

The morphology of M 2-48 has been previously discussed by
V\'azquez et al. (2000) [hereafter Paper I] who found the full extent of its bipolar structure
and indications of the presence of high-velocity outflows from their
monochromatic images. The occurrence  of highly collimated, high velocity
jet-like components in planetary nebulae is well established (e.g. Miranda \&
Solf 1992; L\'opez et al. 1995;  Sahai \&
Trauger 1998;  Miranda, Guerrero \& Torrelles 1999; O'Connor et al. 2000;
Miranda, Guerrero \& Torrelles 2001 and Gon\c{c}alves, Corradi \& Mampaso 2001, etc.). These
bipolar outflows usually show indications of episodic outburst and rotation or
precession of the symmetry axis (e.g. L\'opez et al. 1997).
The origin of such outflows is intimately related to the formation and
shaping of PNe and thus their kinematic characterisation is a key parameter for
the models. Several models have been proposed for the generation of
collimated outflows in PNe, such as pure hydrodynamical models (Icke et al.
1992, Mellema 1995, 1997), models involving stellar rotation
and magnetized winds (R\'{o}\.{z}yczka \& Franco 1996; Garc\'{\i}a-Segura 1997;
Garc\'{\i}a-Segura \& L\'opez 2000), and the influence of  binary system (Soker \&
Livio 1994; Reyes-Ru\'\i z \& L\'opez 1999, Soker 2002).

The bipolar planetary nebula M 2-48 has three main structural components:
(1) an internal bipolar central region (CR); (2) an off-center semi-circular fragmented
shell, containing a region with a bow-shock morphology (this has been
interpreted in Paper I as the possible interaction of the bipolar, collimated
outflow with the shell), and (3) a pair of symmetric knots with a seemingly bow-shock
morphology, located at either side at $\approx 2'$ from  the center (see
Paper I).

In order to analyze the kinematical structure, excitation mechanisms and
chemical abundances of M~2-48, spatially resolved high and low
dispersion  spectroscopy have been now  obtained. We first describe the
observations and the results extracted from the position-velocity (PV) diagrams
in section \ref{Observations}. The kinematical structure and  morphology are
discussed in section \ref{Kinematics}, including a simple model to predict the
orientation of the shocked structures with respect to the plane of the sky. In
section \ref{abundances}, the excitation mechanism and chemical abundances at
different regions of the nebula are derived. Finally, our main conclusions are summarized
in section \ref{Conclusions}.

\begin{figure}
\resizebox{ \hsize}{!}{\includegraphics{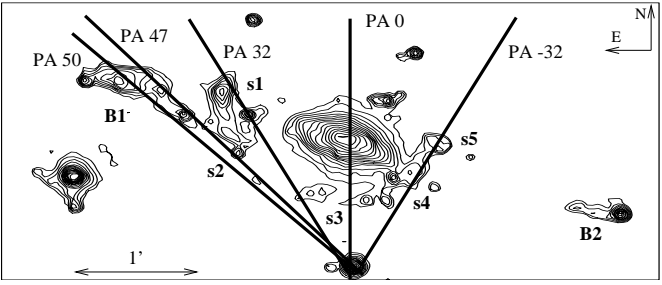}}
\caption{Slit positions of the low-dispersion spectroscopy are marked over the
contour
plot of a [N~II] 6584 \AA \ CCD image of M 2-48 taken from V\'azquez et al. (2000).}
\label{ms2425f1.eps}
\end{figure}

\section{Observations and Results}\label{Observations}

\begin{figure}
\resizebox{ \hsize}{!}{\includegraphics{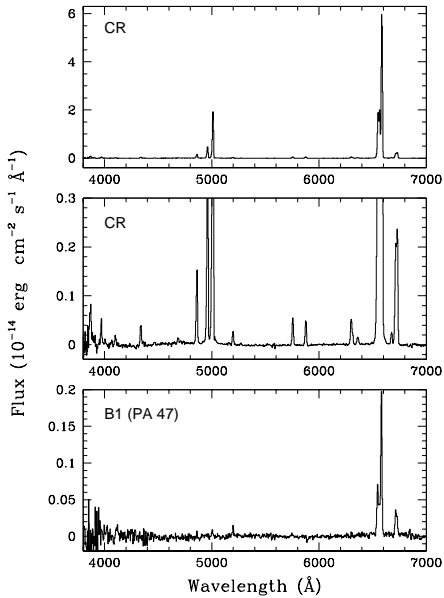}}
\caption{Low-dispersion optical spectra of two different regions of M
2-48.
The central region (CR) is shown expanded along the flux axis in the middle
panel. The bow-shock B1 is shown in the bottom panel.} \label{ms2425f2.eps}
\end{figure}

\noindent
Low-dispersion (LD) spectroscopy was obtained in 1999, July 18, 19 and
20, with the B\&Ch spectrometer combined with the f/7.5 focus of the  2.1-m
OAN telescope at the Sierra de San Pedro M\'artir, B. C., Mexico. A 300 lines/mm grating was
combined with a $220\,\mu$m ($\equiv2.9$\, arcsec) slit width and a Tektronix
CCD of $1024\times1024$ pixel, $24\,\mu$m square pixels ($\equiv 0.84$
arcsec) as detector.
The resulting spectra cover from 3400 to 7500\,{\AA},
with a 12\,\AA \ spectral resolution. The slit was oriented in several position
angles (PA) to cover different regions of the PN. Figure \ref{ms2425f1.eps} shows the slit 
positions over a contour plot of the  [N~II] 6584 {\AA} image (Paper
I).  The LD CCD frames  were trimmed, bias-subtracted, flat-fielded
and 
sky-subtracted with standard techniques using IRAF. Flux calibration
was 
performed using sensitivity functions derived from the standard star
HD\,192281. Table 
\ref{tabla1} presents the log for each of the 1D spectra that have been
extracted and analyzed and Figure \ref{ms2425f2.eps} shows the one-dimensional spectra of CR and B1.

\begin{figure}
\resizebox{ \hsize}{!}{\includegraphics{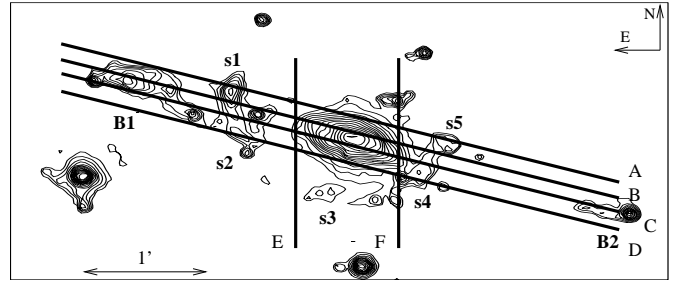}}
\caption{Slit positions of the high-dispersion  spectroscopy (A-F) are marked against
a
contour plot of a [N~II] 6584 \AA \ CCD image of M 2-48 taken from V\'azquez et al.
(2000).}
\label{ms2425f3.eps}
\end{figure}

\begin{table}[!]
\caption[]{Data of the 1D spectra extracted from the low dispersion
spectra.}\label{tabla1}
\begin{flushleft}
\center
\begin{tabular}{lccc}
\hline\hline
\noalign{\smallskip}
&   P.A.& Exposure& Slit length \\
Region& ($^\circ$)& time (s)& (arcsec) \\
\noalign{\smallskip}
\hline
\noalign{\smallskip}
CR& 0& 3600& 25.9 \\
s1& 32& 2400& 11.5 \\
s4& -32& 1800& 8.6 \\
B1& 47& 1800& 20.2 \\
B1& 50& 3000& 4.8 \\
\noalign{\smallskip}
\hline
\noalign{\smallskip}
\end{tabular}
\end{flushleft}
\end{table}

The high-dispersion (HD) spectroscopy was obtained in 1999, June 29 and
30,
with the Manchester Echelle Spectrometer (MES; Meaburn et al. 1984)
combined with the f/7.9 focus at the  2.1-m OAN telescope at the Sierra
de
San Pedro M\'artir, B. C., Mexico. This 
spectrometer
has no cross-dispersion. A {90\,\AA} bandwith filter was used in order
to
isolate the 87th order, which contains the H$\alpha$ and
[N II] $6548, 6584$ \ \AA \ emission lines.

\begin{figure}
\begin{center}
\resizebox{ \hsize}{!}{\includegraphics{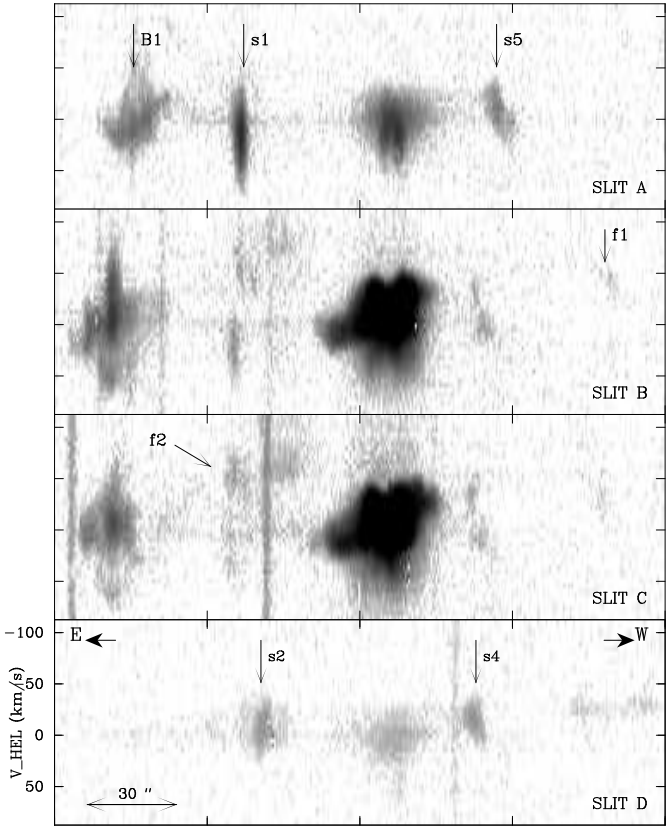}}
\end{center}
\caption[Position-velocity arrays over slits A -- D of M 2-48]{Gray-scale
representation of the
position-velocity arrays of [N II] 6584 \AA \  along slits A -- D.
Labels correspond to the regions indicated in Figure 3.
\label{ms2425f4.eps}}
\end{figure}

\noindent

\begin{figure}[t]
\resizebox{\hsize}{!}{\includegraphics{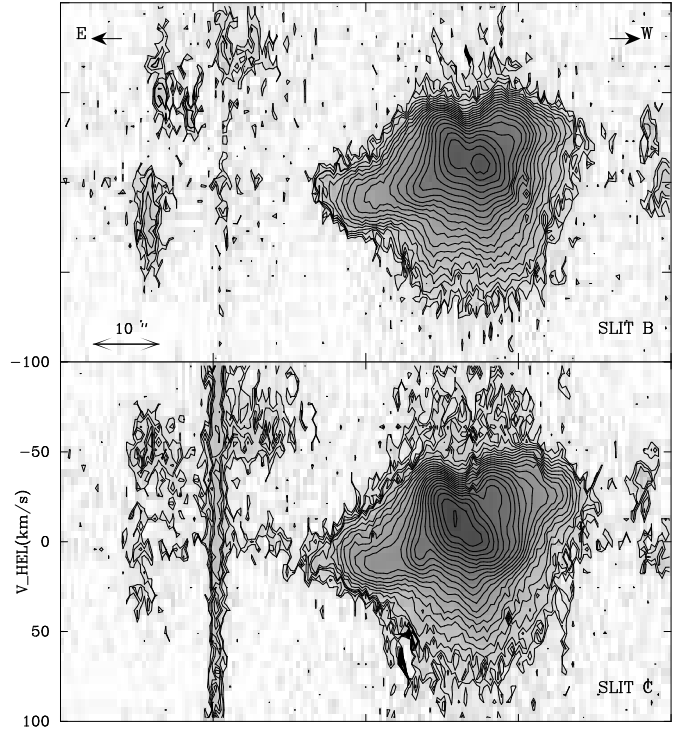}}
\caption{[N II] 6584 \AA \ PV diagrams of the central region of M 2-48
(for
slit positions B and C). The line profiles are shown with a greyscale
and overlayed
logarithmic contours. }
\label{ms2425f5.eps}
\end{figure}

A  Tektronix CCD with $1024\times1024$, 24 $\mu m$ ($\equiv 0.3$ arcsec)
square pixels
was the detector. A 2 $\times$ 2 binning  was employed in both spatial and
spectral dimensions.
 The ``seeing'' varied between 1 and 2 arcsec during the
observations. In Figure \ref{ms2425f3.eps}, the slit positions A-F are
 shown against a
contour plot of  [N~II] $6584$ \,\AA  \ of M~2-48.
The 150 $\mu m \  (\equiv 1.9$ arcsec) wide  slit was oriented at a position
angle
PA=75$^{\circ}$ for slits
A -- D and North-South for slits E --F. The spectral
resolution is
 $10$\,km\,s$^{-1}$. Exposure times of 1800\,s were carried out for
each slit position. Images were reduced in the standard manner using
IRAF.
The spectra were wavelength calibrated with a Th-Ar lamp
to an accuracy of $\pm1$\,km\,s$^{-1}$.

The bidimensional arrays of [N~II] 6584{\,\AA} position-velocity (PV) line
profiles from the HD spectra (slits A-D), are shown in Figure
\ref{ms2425f4.eps}. Figures 5, 6 and 7 show enlargements of the core region
for slits B and C; the line profiles from slits E and F and the line profiles
at the tip of north eastern lobe for slits B and C, respectively.

\section{Morphology and Kinematics}\label{Kinematics}
\noindent
M 2-48 has a  bipolar morphology, as has been described in Paper I (see also the images in Manchado et al. 1996).
From a bright
bipolar core emerge relatively faint, broken filaments that delineate fairly
tight collimated lobes extending over 2 arcmin on either side.
Some of the filaments close to the core seem to trace an incomplete arc that
gives the appearence of being the remnants of a spherical shell.
The PV diagrams  shown in Figures 4 -- 7 reveal the complex kinematic
structure of the main components of M 2-48.
The labels B1, s1, s2, s4 and s5 in the PV arrays in Figure \ref{ms2425f4.eps}
match the location of the regions with the same labels in Figure 3, where we
have followed the nomenclature of Paper I. The spectral element
labeled as (f1) is probably related to the tip of the southwestern lobe.

\begin{figure}
\begin{center}
\resizebox{ \hsize}{!}{\includegraphics{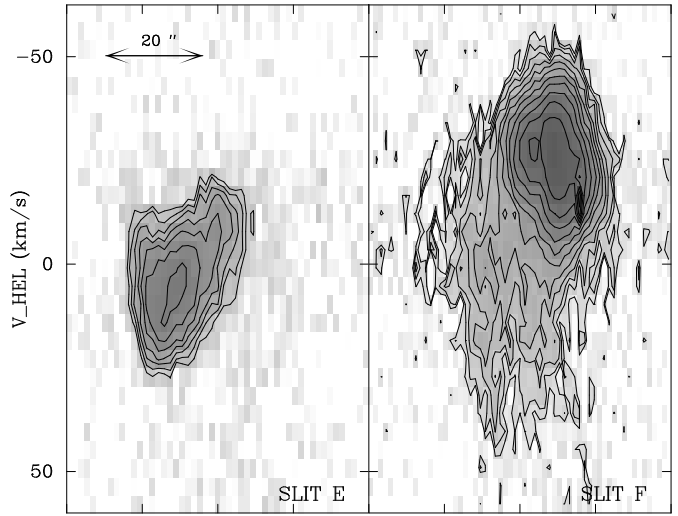}}
\end{center}
\caption{ [N II] 6584 \AA \ PV diagrams of the bipolar lobes of M 2-48
(for
slit positions E and F). The line profiles are shown with a greyscale
and with
logarithmic contours.\label{ms2425f6.eps}}
\end{figure}

\subsection{Central region}\label{central}

\noindent
Line profiles from the bright nebular core are shown in detail in
Figure \ref{ms2425f5.eps}.
The systemic velocity of the core, as defined by the central intensity
maximum of the line profiles shown in  Figure \ref{ms2425f5.eps} lies around
$-15$ km s$^{-1}$. The line profiles  replicate a bipolar
outflow emerging from a bright, compact core,  extending bluewards to -50 km
s$^{-1}$ on the west side and redwards to $+25$ km s$^{-1}$ on the east side.
The line profile from slit C (bottom panel in Fig. 5) shows particularly
extended wings, reaching remarkable values of $\pm 85$ km s$^{-1}$ at FWZI
within the central ($\sim$ 20 arcsec) region.
The bright compact core itself is seen elongated and tilted in the velocity
space, expanding towards negative velocities on the east side ($ -25$ km
s$^{-1}$) and positive velocities on the west side ($+10$ km s$^{-1}$).

\begin{figure}
\begin{center}
\resizebox{ \hsize}{!}{\includegraphics{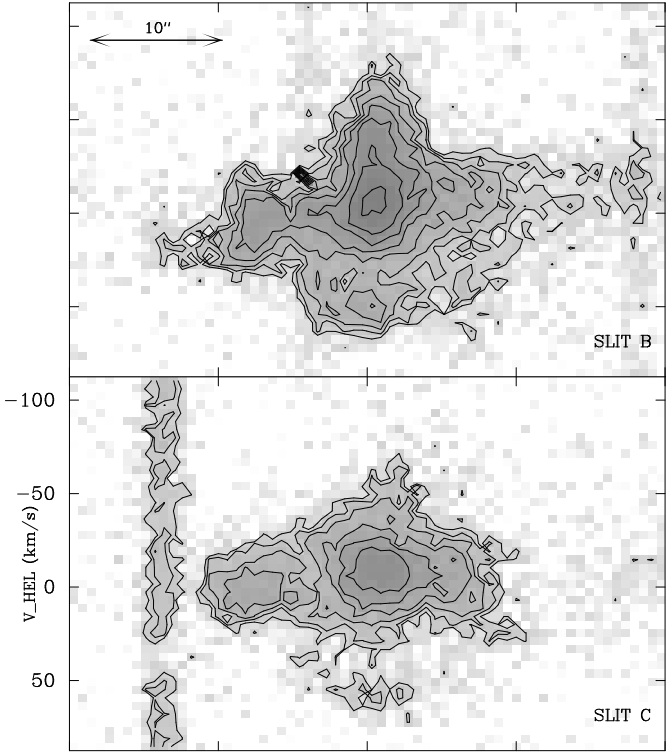}}
\end{center}
\caption{ [N II] 6584 \AA \ PV diagrams of the bow-shock B1 of M 2-48
(for
slit positions B and C). The line profiles are shown with a greyscale
and with
logarithmic contours.\label{ms2425f7.eps}}
\end{figure}

\subsection{The arc-like structure}\label{arcs}
\noindent

One of the curious features of M 2-48 is a group of several
knots around the central bipolar region that seem to form
an arc-like structure (Paper I).
These knots (s1-s5 in Figure \ref{ms2425f1.eps}) were  interpreted  as possible
segments of an expanding semi-circular shell. Some
slits were obtained with the purpose of intersecting 
this possible expanding shell and clarify their nature. 
The line profiles shown in Figure \ref{ms2425f6.eps}, are found to follow
the velocity pattern of the bipolar outflow emerging from the extended regions
of the core, with redshifted velocities on the east side ($\approx$ $+10$  km
s$^{-1}$) and blueshifted velocities on the west side ($\approx$ $-25$  km
s$^{-1}$). No indications of a velocity ellipse that could hint to the presence
of an expanding shell are detected, at least with the present spectral
resolution.

\subsection{Bow-shock s1}\label{lobes}
\noindent
The eastern lobe is the brighter one, with several high-speed knots
with bow-shock looking morphologies being detected in the HD spectra, namely
B1, s1 and s2. On the western lobe we detect and
identify knots s4 and s5, and the feature f1 (see Figures \ref{ms2425f3.eps}
and \ref{ms2425f4.eps}).

\bigskip

\begin{figure}[t]
\begin{center}
\resizebox{ \hsize}{!}{\includegraphics{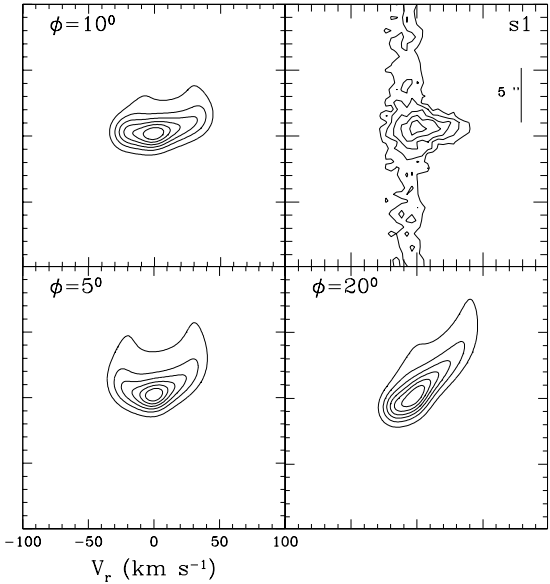}}
\end{center}
\caption{Comparison of the position-velocity diagram of the  s1 bow-shock
 and ``3/2-D'' bowshock model fits at different orientations pointing away from the plane of the sky. The best fit is at $\phi = 10 ^\circ$ .\label{ms2425f8.eps}}
\end{figure}

A high-velocity element, labeled s1 in Figure 4,
coincident with the position of a bright knot on the
eastern side and near a portion of the arc-like structure, shows an apparent
bow- shock morphology and kinematical structure (see Figures \ref{ms2425f3.eps}
and \ref{ms2425f8.eps}, bottom right and the image in Manchado et al. 1996).
It is of interest to estimate the orientation angle of the bow-shock
associated with knot s1 with respect to the  plane of the sky, $\phi$. To do that, we 
have used a  ``3/2-D'' model to predict the position-velocity diagram of a 
bow-shock at different angles with respect to the plane of the sky.
 Raga \&
Noriega-Crespo (1993) found that the H$\alpha$ emission per unit time
and area is approximately given by~:

\begin{equation}
I_{\rm H\alpha} \propto \left( v_{shock} \right)^{\gamma_{\rm
H\alpha}}\,,
\label{m24821}
\end{equation}
\noindent
where $v_{shock}$ is the shock velocity and $\gamma_{\rm H\alpha}$ is a
constant dependent on the shock velocity regime.

If we assume that 
the jet is moving along the  $ z$-axis, the shape of the bow-shock can be
parametrized
by~:

\begin{equation}
r \propto z^p\,, \label{m24822}
\end{equation}
where $p$ determinates the aperture of the bow-shock. We have used a value of $p=2$.

Thus, the input parameter for the model  essentially is the shock velocity.
This velocity can be estimated from the PV diagrams obtained from  the
long-slit spectroscopy (Figure \ref{ms2425f4.eps}).

In addition, we use a value of $\gamma_{\rm H\alpha}= 3.49$
corresponding to the 20.0--81.3 km s$^{-1}$
range of velocities proposed by Raga \& Noriega-Crespo (1993).

We can calculate the H$\alpha$ emission, and predict
the position-velocity diagram at different orientations with respect
to the sky plane, as
shown in Figure \ref{ms2425f8.eps}. From this comparison
we can
establish that s1 is inclined at an angle $\sim 10^{\circ}$  with
respect
to the plane of the sky.

This low inclination angle indicates that the radial velocity of the
outflow 
producing the s1 bow-shock  can be a factor of about 6 larger than the 
velocities derived from the PV diagrams.

 We have used the radial velocity of the peak emission of s1,$~$ 15 km s$^{-1}$, as the expansion velocity projected to the line of sight. Using the inclination estimated for s1, $\phi \sim10^{\circ}$, we can determine the tangential expansion velocity. Finally, taking into account the projected linear separation between s1 and the central region we can estimate a dynamical time of $\tau_{\rm s1} \approx 3 \left[ d/kpc \right] \times 10^3$ yr.

In addition, a peculiar feature labelled f2 can be
distinguished in the PV diagram of slit C (Figure \ref{ms2425f4.eps}),
 with a constant 
heliocentric velocity of about  $-$60 km s$^{-1}$. This feature 
seems to be spatially coincident with the eastern arc, 
corresponding to the diffuse material between knots s1 and s2.

\subsection{External bow-shock B1}\label{bow-shock}
\noindent

\begin{figure}[t]
\begin{center}
\resizebox{ \hsize}{!}{\includegraphics{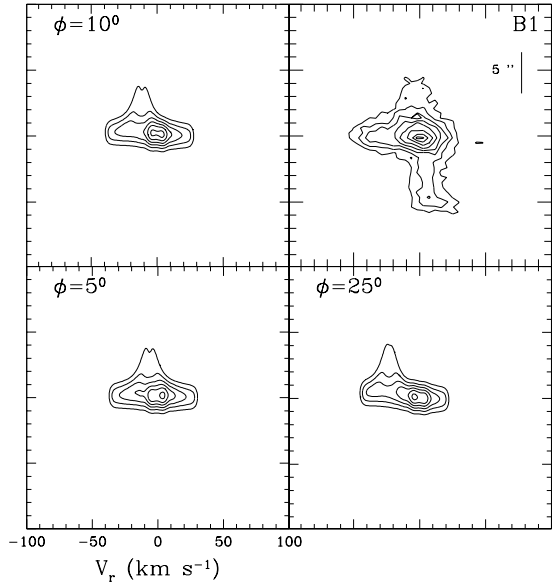}}
\end{center}
\caption{Comparison of the position-velocity diagram of the  B1 bow-shock
 and ``3/2-D'' bowshock model fits at different orientations pointing inwards to  the plane of the sky. The best fit is at $\phi = 10 ^\circ$. \label{ms2425f9.eps}}
\end{figure}

\noindent
In the [N~II] 6584 \AA \ direct images (see Paper I), it 
is possible to see two symmetrical structures with a bow-shock
morphology. The
eastern knot, B1, which is brighter than the western one, B2,  is blueshifted
and has    
 a line width of  $\approx 80$  km s$^{-1}$.
The PV diagrams of this bow-shock are shown in Figure
\ref{ms2425f7.eps},
where we can see two different zones of B1  covered by the slits B and
C. Slit B slightly covers the center of the
bow-shock
(larger line width), and C covers the wings (smaller line width).

\noindent
It is possible to use the ``3/2-D'' model  described in section
\ref{arcs} to predict the PV diagram of this bow-shock. 
From the comparison of the observations and the numerical model of the
bow-shock B1 for different orientations (see Figure \ref{ms2425f9.eps}), we can estimate that
B1 is
inclined at an angle of $\sim 10^{\circ}$ pointing inwards to the plane
of the
sky. 
  The angle with respect to the  plane of the sky  suggests that the radial 
velocity
of the
bow-shock B1 can be a factor of $\approx 6$ greater than the velocities
derived
from  the PV diagrams. In order to establish a sequence of events,  using the same procedure as for s1 (see section \ref{lobes}), we estimate the
dynamical
age of this bow-shock to be $\tau_{\rm B1} \approx 5.3 \left[ d/kpc \right] \times 10^3$ yr.
This is consistent with  the idea that the bow-shocks B1 and s1 correspond to
different bipolar ejections and probably with different axis orientation. This
suggests a
change in the orientation  of the symmetry axis of the succesive outflows. 
In this way, a  previous ejection of material  would form the expanding
arcs
interacting  with a bipolar outflow and the most external bow-shocks B1
and
B2, and a later high-velocity bipolar outflow with different orientation would form the s1
bow-shock. 
 
We have also noted that the symmetry axis of the most internal
bipolar
region is not coincident with either of these two high-velocity bipolar
outflows.

\begin{table}
\caption[Dereddened line intensity ratios of M 2-48]{Dereddened line
intensity
ratios.}
\begin{flushleft}
\begin{tabular}{lcccc}
\noalign{A - Bright regions}
\noalign{\smallskip}
\hline\hline
\noalign{\smallskip}
Line& f($\lambda$)& CR& s1& B1(47$^\circ$) \\
\noalign{\smallskip}
\hline
\noalign{\smallskip}
3870 He I+[Ne III]& +0.228& 107& ...&  ... \\
3969 H7+[Ne III]& +0.204& 51& ...&  ... \\
4007 [Fe III]& +0.195& 9& ...& ... \\
4101 H$\delta$& +0.172& 21& ...&  ... \\
4340 H$\gamma$& +0.129& 50& ...&  ... \\
4363 [O III]& +0.124& 9& ...&  ... \\
4471 He I& +0.095& 8& ...& ... \\
4686 He II& +0.048& 10& ...& ... \\
4711 [Ar IV]& +0.036& 5& ...& ... \\
4740 [Ar IV]& +0.032& 3& ...& ... \\
4861 H$\beta$& 0.0& 100& 100& 100 \\
4959 [O III]& $-$0.023& 278&  ...& ... \\
5007 [O III]& $-$0.024& 965& 103&172 \\
5152 [Fe III]& $-$0.064& 2& ...& ... \\
5198 [N I]& $-$0.074& 15& ...& 121 \\
5270 [Fe III]& $-$0.087& 2& ...& ... \\
5410  He II& $-$0.116& 1&  ...& ... \\
5755 [N II]& $-$0.191& 16& 32&  ... \\
5876  He I& $-$0.216& 14& ...& ... \\
6300 [O I]+[S III]&  $-$0.285& 14& 72&  ... \\
6364 [O I]&  $-$0.294& 4& 34& ...\\
6548 [N II]& $-$0.321& 285& 214& 481 \\
6563 H$\alpha$& $-$0.323& 286& 275& 286 \\
6584 [N II]& $-$0.326& 860& 622& 1340 \\
6678  He I& $-$0.338& 4&  ...& ... \\
6717 [S II]& $-$0.343& 29& 121& 192\\
6731 [S II]& $-$0.345& 35& 95&139 \\
7065  He I& $-$0.383& 4&  ...& ... \\
7135 [Ar III]& $-$0.391& 25& ...& ... \\
7330 [O II]&  $-$0.410& 8&  ...& ... \\
\noalign{\smallskip}
$C$(H$\beta$)& & 1.88& 1.82& 1.47 \\
$I$(H$\beta$)$^{\rm 1}$& & 171& 6.0& 2.1\\
$N_{\rm e}$( [S II]) & & 1260& 180&$<$100 \\
$T_{\rm e}$( [O III])& & 10850& ...& ...\\
$T_{\rm e}$( [N II])& & 10700& 20100& ...\\
\hline
\noalign{\smallskip}
\multicolumn{5}{l}{$^{\rm 1}$ in units of 10$^{-14}$ erg cm$^{-2}$
s$^{-1}$.}\\
\label{tabla2}
\end{tabular}

\end{flushleft}
\begin{flushleft}
\begin{tabular}{lcc}
\noalign{B - Faint regions}
\noalign{\smallskip}
\hline\hline
\noalign{\smallskip}
Line& s4 (-32$^\circ$) & B1(50$^\circ$) \\
\noalign{\smallskip}
\hline
\noalign{\smallskip}
6548  [N II]&  ...& 122 \\
6563 H$\alpha$&  100&  100\\
6584 [N II]&  82& 410 \\
6717+31  [S II]&  55& 152\\
\noalign{\smallskip}
$F$(H$\alpha$)$_{\rm obs}^{\rm 1}$& 1.1& 0.6\\
\hline
\noalign{\smallskip}
\multicolumn{3}{l}{$^{\rm 2}$ in units of 10$^{-15}$ erg cm$^{-2}$
s$^{-1}$.}\\
\noalign{\smallskip}
\end{tabular}
\end{flushleft}
\label{tabla2}
\end{table}

In addition to the main features described in previous sections, we can
see
other interesting features in the PV diagrams obtained with the HD
spectroscopy. 
In Figure \ref{ms2425f4.eps} we can distinguish a 
very weak feature labelled  f1 in slits B and C. This is a component
with an
heliocentric 
velocity of $\approx$ $-50$  km s$^{-1}$. This emission could 
correspond to the wing of
the western bipolar  lobe outflow. Observations at other slit positions or
longer
exposure times would be necessary to resolve in more detail the
kinematical
behavior of this feature.

\section{Line intensity ratios and chemical
abundances}\label{abundances}
\noindent

\begin{figure}
\resizebox{ \hsize}{!}{\includegraphics{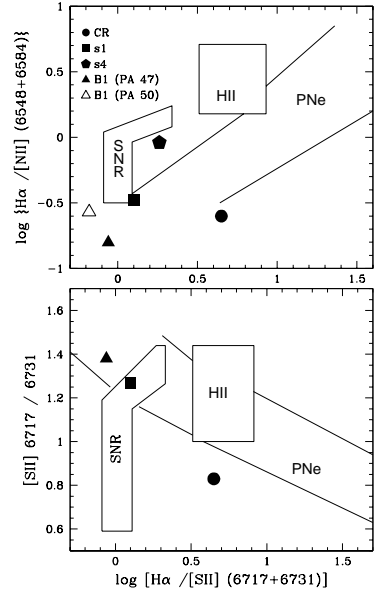}}
\caption[Diagnostic diagrams showing the emission line ratios of the
different
regions observed in M 2-48]{Diagnostic diagrams showing the emission
line
ratios of the different regions observed in M 2-48. Top: log
H$\alpha$/[SII]
vs. log H$\alpha$/[NII]. Bottom: log H$\alpha$/[SII] vs. the electron
density
indicator [SII] 6717/6731 (adapted from Sabbadin et al. 1977)}.
\label{ms2425f10.eps}
\end{figure}

\noindent 
\begin{table}
\caption[Chemical abundances in the central region of M 2-48]{Chemical
abundances in the central region}
\begin{flushleft}
\center
\begin{tabular}{lc}
\hline\hline
\noalign{\smallskip}
12+log O$^+$/H$^+$& 7.85 \\
12+log O$^{++}$/H$^+$& 8.43 \\
12+log O/H& 8.55 \\
log N$^+$/O$^+$& +0.30 \\
12+log Ne$^{++}$/H$^+$& 7.75\\
12+log S$^+$/$^+$& 6.16\\
12+log Ar$^{++}$/H$^+$& 6.30\\
12+log Ar$^{3+}$/H$^+$& 5.77\\
\noalign{\smallskip}
He$^+$/H$^+$ (4471)& 0.127\\
He$^+$/H$^+$ (5876)& 0.098\\
He$^+$/H$^+$ (6678)& 0.096 \\
He$^+$/H$^+$ (7065)& 0.163 \\
$<$He$^+$/H$^+$$>$& 0.100 \\
He$^{++}$/H$^+$& 0.008 \\
He/H& 0.108 \\
\hline
\end{tabular}
\end{flushleft}
\label{tabla3}
\end{table}

As noted in section \ref{Observations}, five regions were
selected
from the slit positions observed with low resolution spectroscopy (see
Figure
\ref{ms2425f1.eps}), covering the four different morphological zones pointed 
out
in Paper I. In Figure \ref{ms2425f2.eps}, the flux calibrated spectra of
two
different zones are shown. The top and middle panels correspond to
different
scales (high and low intensities, respectively) of the spectrum of the
central
region. The bottom panel corresponds to the spectrum of B1 (PA
47$^\circ$). 

The extinction coefficient, C(H$\beta$), was derived for three regions:
Central Region (CR), s1
and B1 (PA 47$^\circ$) from the comparison of observed and theoretical
H$\alpha$/H$\beta$ ratios. We have used the calculations of Storey \&
Hummer
(1995) assuming case B and an iteration procedure adopting finally the
electron densities ($N_e$) derived for each particular zone and the electron
temperature ($T_e$) obtained for CR ($T_e$=10800 K) for all of the regions. The 
reddening law of Seaton (1979) has been used. Dereddened line
intensity
ratios with respect to H$\beta$ as well as the values of  C(H$\beta$) for these
 regions
are
presented in Table \ref{tabla2}. The uncertainties in the line fluxes
are of
the order of 5\% for the brighter lines and about 20-30\% for the
weaker
ones. Emission line fluxes of blended lines such as the H$\alpha$+[N~II] nebular
lines
and the [S~II] doublet have been obtained from Gaussian fitting of the
line
profiles using the Starlink DIPSO package (Howarth \& Murray 1990). The
dereddened, integrated H$\beta$ flux of each region is also included in
Table
\ref{tabla2}. 

\begin{figure}
\resizebox{ \hsize}{!}{\includegraphics{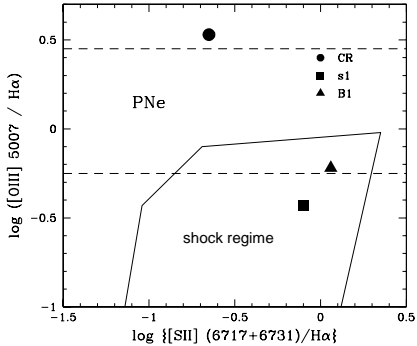}}
\caption[Diagnostic diagram representing ratios of  emission lines of
the
different regions observed in M 2-48 ]{ Diagnostic diagram representing
log
[SII]/H$\alpha$ vs. log [OIII]/H$\alpha$ of the different regions
observed in
M 2-48 (adapted from Phillips \& Cuesta 1999)}
\label{ms2425f11.eps}
\end{figure}

$N_e$ and $T_e$ (see Table \ref{tabla2}) have been
derived
following standard methods of collisionally excited line ratios using
the
five-level program for the analysis of emission-line nebulae of Shaw \&
Dufour
(1995). There seems to be a gradient in $N_e$, with 
parameter
decreasing values  towards the external zones. This is consistent with the
expansion of
stellar ejecta as the distance towards the nucleus increases. $T_e$
 derived from [O~III] and [N~II] lines for CR are almost
coincident, indicating that there is no strong temperature
stratification in
the bright bipolar nebular core of the nebula. The $T_e$ derived
for s1
seems too high to be accounted for 
by pure photoionization, confirming  contamination by
shock
emission  in this zone, which morphologically and kinematically
resembles a
bow-shock (see section \ref{lobes}).

The spectra of the other two zones: s4 and B1 (PA 50$^\circ$), are so
faint
that H$\beta$ is not measured and therefore it is not possible to derive
the
extinction coefficient for these regions. The emission line ratios with
respect H$\beta$ and the uncorrected H$\alpha$ flux for these regions 
are included in Table \ref{tabla2}. 

We have used several diagnostic diagrams in order to evaluate which is 
 the dominant
excitation
mechanism prevalent in these regions. In Figure \ref{ms2425f10.eps} we
show
two diagrams involving the locus of several line ratios:
H$\alpha$/[N~II],
H$\alpha$/[S~II], and [S~II] 6717/6731 for H~II regions, PNe and
supernova
remnants (adapted from Sabbadin et al. 1977). These diagrams 
clearly indicate that CR is mainly radiatively excited. However, the positions 
of s1 and B1 are not so definite, in the sense that some shock contribution
is
present. The fact that the positions of B1 and s1 are just below the 
locus of the SNRs in the upper pannel of Figure \ref{ms2425f10.eps} could be also due to the high N enrichment of the nebula (see below).

In Figure  \ref{ms2425f11.eps} we show a diagram adapted from Phillips
\&
Cuesta (1999) involving [O~III]/H$\alpha$ and [S~II]/H$\alpha$. This
diagram
includes the locus of the observed emission line ratios for a large number
of
radiatively excited PNe and predictions from  plane-parallel and bow-shock
models
by Hartigan et al. (1987) and Shull \& McKee (1979). The position of the 
different regions of M 2-48 in this diagram confirms that CR is
radiatively 
excited and that  shock excitation is  contributing to the 
spectra of s1 and B1.  Assuming purely shock  excitation for these
zones
and comparing their line intensities with the models by Hartigan et al.
(1987)
we find the best agreement for models with shock velocities of 60 and
130 km
s$^{-1}$, respectively, in agreement with the values obtained in previous 
sections.
 
Ionic abundances of the CR are shown in Table \ref{tabla3} and have been
derived using the emission line ratios given in Table \ref{tabla2},  the
code
of Shaw \& Dufour (1995), and assuming T$_e$ = 10\,800 K and N$_e$ = 1260
cm$^{-3}$. To derive the He$^+$/H$^+$ and He$^{++}$/H$^+$ ratios we have
used
the effective recombination coefficients given by P\'equignot et al. 
(1991)
for all the lines involved. The He$^+$ abundances obtained  from each 
individual line have  been corrected for collisional contributions
following
Kingdon \& Ferland (1995). The final,  adopted value of the He$^+$
abundance
is the average of the values obtained for the different individual lines
excluding HeI 7065 \AA, which suffers from the largest collisional
effects. 

The total O abundance has been determined from the derived abundance of
O$^+$ and
O$^{++}$ combined with the ionization correction factor (ICF) for the
presence
of O$^{3+}$ proposed by Kingsburgh \& Barlow (1994). The total He/H
ratio has
been obtained simply by adding the $<$He$^+$/H$^+>$ and He$^{++}$/H$^+$ ratios. 

The log N$^+$/O$^+$=+0.30 obtained for the nebula clearly indicates that
it is
a nitrogen-enriched object. Taking into account the chemical
classification
proposed by Peimbert \& Torres-Peimbert (1987), M 2-48 should be
classified as
a Type I PN (log N/O $\geq$ $-$0.30 and/or He/H $\geq$ 0.125), although
it is
not a helium-rich object. The O/H ratio is slightly lower than the
average of
the Galactic Type I PNe (12+log O/H = 8.65, Kingsburgh \& Barlow 1994).

\section{Conclusions}\label{Conclusions}
\noindent
The kinematic mapping of M 2-48 has revealed a complex bipolar structure characterised by a series of shocked regions extending accross the major axis of the nebula and reaching the tips found in paper I. Indications that the bipolar outflow has suffered some changes in direction have also been detected. 3/2 D shock models of the line profiles are consistent with the main bipolar outflow having a small $\pm$ 10$^{\circ}$ angle with respect to the plane of the sky, thus making the real outflow velocity some six times larger than the observed one.
The line fluxes derived from the low dispersion spectrum in the central region show that M 2-48 is nitrogen enriched; and should be classified as  type I.

\acknowledgements

We would like to thank the referee, You Hua Chu, for useful comments that improved our paper. L\'opez-Mart\'\i n is in grateful receipt of a graduate scholarship from DGEP-UNAM (M\'exico). L\'opez-Mart\'\i n and Raga acknowledge support
from the CONACyT grant 34566-E. J.A. L\'opez ackowledges support from CONACYT and DGAPA (UNAM) through projects 32214-E and IN114199. V\'azquez was partially
supported by the CONACyT grant I32815E. Miranda and Torrelles acknowledge 
support from DGESIC grant PB98-0670-C02, and from Junta de Andaluc\'\i a 
(Spain).Olgu\'\i n is in grateful receipt of a graduate scholarship from 
DGAPA-UNAM (M\'exico). We thank R. Cook and the SPM staff for help during the observations.

\def\bibname{References}
--------------------------------------------------------------------

\clearpage

\end{document}